%% file: main.tex
\documentclass[a4paper,USenglish,cleveref,autoref,thm-restate]{lipics-v2021}

\hideLIPIcs

\usepackage{tikz}
\usetikzlibrary{arrows.meta, positioning}
\usepackage{booktabs}
\usepackage{longtable}
\usepackage{url}
\usepackage{colortbl}
\usepackage{float}

\let\oldbibliography\thebibliography
\renewcommand{\thebibliography}[1]{%
  \oldbibliography{#1}%
  \setlength{\itemsep}{0pt plus 1pt}%
  \setlength{\parsep}{0pt plus 1pt}%
}

\newif\ifLIPIcsLayout
\LIPIcsLayouttrue

\newif\ifFigDistributions
\FigDistributionstrue

\definecolor{t1col}{HTML}{2d5f8a}
\definecolor{t2col}{HTML}{3a7d5c}
\definecolor{t3col}{HTML}{7b4ea0}
\definecolor{t4col}{HTML}{b84c1a}
\definecolor{t5col}{HTML}{3a5a8c}
\definecolor{structgrey}{HTML}{aaaaaa}
\definecolor{rowshade}{HTML}{f7f4ef}
\definecolor{sectionshade}{HTML}{e8e8e8}

\newcommand{\dimtag}[2]{\colorbox{#1}{\textcolor{white}{\textbf{\small\texttt{#2}}}}}

\title{Beyond the Grave: An Empirical Study of Dormancy and Revival in Scientific Open-Source Software}
\titlerunning{Dormancy and Revival in Scientific Open-Source Software}

\author{Addi Malviya Thakur}{University of Tennessee, Knoxville, TN, USA \and Oak Ridge National Laboratory, Oak Ridge, TN, USA}{amalviya@vols.utk.edu}{}{}

\author{Bogdan Vasilescu}{Carnegie Mellon University, Pittsburgh, PA, USA}{vasilescu@cmu.edu}{}{}

\author{Audris Mockus}{University of Tennessee, Knoxville, TN, USA}{audris@utk.edu}{}{}

\authorrunning{A. Malviya Thakur, B. Vasilescu, and A. Mockus}

\Copyright{Addi Malviya Thakur, Bogdan Vasilescu, and Audris Mockus}



\category{Empirical Track}

\relatedversion{}

\supplement{\url{https://anonymous.4open.science/r/ESEM2026ReplicationPackage-598E}}

\funding{}

\begin{document}
\nolinenumbers

\maketitle

\begin{abstract}
\input{ESEM_version/v67/1_ESEM_Edits_Abstract_v67}
\end{abstract}

\input{ESEM_version/v67/2_ESEM_Edits_Introduction_v67}
\input{ESEM_version/v67/3_ESEM_Edits_RelatedWork_v67}

\input{ESEM_version/v67/4_ESEM_Edits_Methodology_v67}
\input{ESEM_version/v67/5_ESEM_Edits_Results_Section_v67}
\input{ESEM_version/v67/6_ESEM_Edits_Discussion_v67}
\input{ESEM_version/v67/7_ESEM_Edits_Threats_v67}
\input{ESEM_version/v67/8_ESEM_Edits_Conclusion_v67}

\input{ESEM_version/v67/9_ESEM_Edits_DataAvailability_v67}

\section*{Acknowledgements}

We thank the cohort of graduate and advanced-undergraduate
analyst-coders whose careful and structured coding of 750 scientific
OSS repositories made this study possible. Their commitment to
evidence-anchored qualitative coding under the two-phase peer-validation
and supervised-adjudication protocol made every reported result possible.

We also thank the maintainers and contributors of the \textsc{SciCat} corpus.
The systematic identification and abandonment classification of 18{,}247
scientific OSS repositories provided the empirical foundation that
enabled our stratified sampling and downstream analyses.

This study depends on a broader open-source ecosystem we gratefully
acknowledge: Python, pandas, NumPy, SciPy, scikit-learn, statsmodels,
matplotlib, and Jupyter for the analytic pipeline; Git and GitHub's
public APIs for repository observation; \LaTeX, TikZ, and the LIPIcs
document class for manuscript preparation.

\bibliographystyle{plainurl}
\bibliography{ESEM_version/v67/esem_cite}

\end{document}

%% file: ESEM_version/v67/1_ESEM_Edits_Abstract_v67.tex
\textbf{Background.} Inactivity-based thresholds are a standard approach for
classifying scientific open-source software (OSS) as abandoned, but they
cannot distinguish permanent abandonment from temporary dormancy.
Threshold choice itself is unstable: moving the inactivity cutoff from
1 month to 36 months changes the number of projects classified as
abandoned in the \textsc{SciCat} corpus from 18{,}030 to 8{,}010.
Whether, how, and how durably such projects revive has not been
systematically studied.
\textbf{Aims.} We characterize what happens after apparent
abandonment in scientific OSS: the causes of dormancy, the
mechanisms of revival, the durability of recovery, and the alignment
between lifecycle archetype  (the overall temporal pattern of
activity, dormancy, and post-revival trajectory) and revival sustainability outcome.
\textbf{Method.} From the \textsc{SciCat} corpus of 18,247 scientific OSS
repositories~\cite{malviyathakur2025scicat}, we identify 2,984
dormant-revived candidates and field-code a stratified sample of 750
projects under a structured qualitative rubric (informed
by Strauss and Corbin~\cite{strauss1990grounded} for the inductive
category-derivation phase). 75
graduate and advanced-undergraduate student analyst-coders (from a
pool of 90) perform field-level coding with two-phase peer validation and supervised
adjudication; post-adjudication $\kappa$ ranges from 0.779 to 0.857.
A deterministic rule-based
content-analysis classifier~\cite{krippendorff2004content,defranco2017content}
produces the T1--T5 categorical labels
from the coded dataset; we publish the classifier's full
operationalization in the replication package and
analyze T1--T5 associations with $\chi^2$, Cram\'{e}r's $V$, and
Kruskal--Wallis tests.
\textbf{Results.} We report sample estimates under our
visibility-weighted sampling design (median 170 GitHub stars vs.\ 9
for the parent \textsc{SciCat} abandoned corpus), not population
estimates. We cannot resolve dormancy cause from repository evidence
for 52.5\% of projects (a measurement-validity ceiling on
repository-only attribution); among the resolvable subset,
feature/milestone freeze outnumbers research-output completion 5.4:1
(156 vs.\ 29 projects), complicating the widespread assumption that
scientific OSS dormancy is primarily publication driven. The
two dominant sustainability outcomes are Sustained Recovery and
Recovered-Then-Declined (jointly 59.5\%); aggregated non-sustained
recovery outnumbers sustained recovery, rank-stable across imputation
choice. 11.5\% of apparent revivals are bot-only or single-spike
artifacts. In this operationalization, the lifecycle archetype shows a
stronger revival sustainability association (medium effect on the
subset of archetypes whose definitions do not predetermine a
sustainability outcome) than
revival mechanism or work type: revival mechanism is non-significant
and work type weak and inconclusive.
\textbf{Conclusions.} A fixed inactivity threshold appears insufficient to reliably classify
scientific OSS abandonment; revival outcome is associated with inactivity
gap, lifecycle archetype, and contributor continuity. We argue against
relying on binary inactivity-based classification alone and motivate
multi-signal indicators grounded in these three dimensions.

%% file: ESEM_version/v67/2_ESEM_Edits_Introduction_v67.tex
\section{Introduction}
\label{sec:intro}
\ifLIPIcsLayout\vspace{-6pt}\fi

Scientific open-source software occupies a peculiar niche in the
broader OSS landscape: it sustains research communities, encodes
hard-won domain expertise, and yet may be fragile in ways that
general-purpose OSS is not. A researcher who depends on a niche
simulation package
cannot easily switch to a competing implementation; a funder making
portfolio decisions about which research-software efforts to renew needs
to know whether a quiet repository is dead or simply between papers or funding; a
tool that monitors scientific software health and flags projects as
abandoned will mislead its users if its definition of \emph{abandoned}
is wrong. In each of these settings, the prevailing operationalization is
the same: classifying a project as abandoned after several months without
a commit~\cite{avelino2019abandonment, coelho2018modern, malviyathakur2025scicat}.
The cost of getting it wrong differs by
setting, but is real in each.

Malviya Thakur et al.\ introduce \textsc{SciCat}, a corpus of 18,247
scientific OSS repositories, and classify 11,631 as abandoned under a
six-month inactivity threshold~\cite{malviyathakur2025scicat}.
Using their data, we find that, of the 18,030 projects with sufficient
commit-history timestamps for threshold evaluation, changing the
inactivity threshold from one month to 36 months reduces the count
classified as abandoned from 18,030 to 8,010, a more than 50\% drop
driven entirely by the choice of threshold rather than any change in
project state.
The reverse direction shows the same instability: 2,984 of the
projects active in the past month, and 2,285 of those active in the
past 12 months, each had at least one prior gap exceeding the
corresponding inactivity threshold, and would have been classified
as abandoned at some earlier snapshot.
Such high uncertainty raises a measurement question that downstream
analyses inherit. Motivated by this, we focus on the subset of
\textsc{SciCat} projects classified as abandoned yet showing renewed
commit activity (\emph{dormant-revived} or \emph{zombie} projects),
and field-code a stratified sample of 750 projects using a structured
field-coding rubric applied by 75 graduate and advanced-undergraduate
student analyst-coders working under supervision. We then process the coded dataset
with a deterministic rule-based content-analysis
classifier (§\ref{sec:taxonomy_construction}) that produces a
five-dimensional classification of the dormancy-revival lifecycle:
dormancy cause (T1), revival mechanism (T2), nature of revival work
(T3), revival sustainability (T4), and lifecycle archetype (T5), 
the
overall temporal pattern of activity, dormancy, and post-revival
trajectory.
This five-dimensional view motivates a lifecycle treatment of dormancy
and revival rather than a binary active/abandoned label; quantitative
magnitudes appear in §\ref{sec:results}.
The five research questions follow the dormancy-revival lifecycle in
sequence. RQ1--RQ2 describe history, RQ3 measures outcome, RQ4 names
the pattern, and RQ5 quantifies the pattern--outcome relationship.
\begin{itemize}
  \item \textbf{RQ1 (Dormancy event):} To what extent can the cause of
  extended dormancy in scientific OSS be inferred from
  repository-visible evidence, and among resolvable cases, what causes
  dominate?
  \item \textbf{RQ2 (Revival event):} What mechanisms trigger revival
  in dormant scientific OSS projects, and what types of work
  characterize recovery activity?
  \item \textbf{RQ3 (Revival outcome):} How often is apparent
  abandonment followed by meaningful recovery, and how sustainable is
  that recovery?
  \item \textbf{RQ4 (Holistic trajectory):} What lifecycle archetypes
  characterize dormant-revived scientific OSS projects?
  \item \textbf{RQ5 (Trajectory--outcome alignment):} How do lifecycle
  archetypes align with revival sustainability outcomes?
\end{itemize}

\noindent This paper makes six contributions.
\textbf{First}, a \emph{threshold-instability finding}: on the same
\textsc{SciCat} corpus, varying the inactivity threshold from one month
to 36 months reduces the count classified as abandoned from 18,030 to
8,010, a more than 50\% swing driven entirely by threshold choice
rather than any change in project state (reproducible from the released
candidate-extraction scripts).
\textbf{Second}, a \emph{measurement-validity ceiling for repository-only
dormancy-cause attribution} (RQ1, §\ref{sec:rq1}): 52.5\% of codeable
projects fall outside resolvable cause categories, quantifying a bound
on what automated analysis of repository signals alone can achieve.
\textbf{Third}, \emph{feature/milestone freeze dominates
research-output completion 5.4:1} among resolvable causes (RQ1, §\ref{sec:rq1}), complicating the publication-driven model of
scientific software.
\textbf{Fourth}, \emph{revival mechanism and work type are weak or
inconclusive} sustainability signals (RQ2, §\ref{sec:rq2}).
\textbf{Fifth}, a \emph{third post-abandonment regime}
(RQ3, §\ref{sec:rq3}): in our dormant-revived sample, apparent
abandonment is often temporary but durable recovery is not the
dominant outcome; Recovered-Then-Declined (31.6\%,
$n=237$) is the modal outcome, and 11.5\% of apparent revivals are
bot-only or single-spike artefacts.
\textbf{Sixth}, \emph{lifecycle archetype is a stronger sustainability
signal than mechanism or work type} in this operationalization
(RQ4, RQ5, §\ref{sec:rq5}): nine archetypes whose association with
outcome exceeds that of mechanism or work type on a
structurally-independent subset.
\textbf{Finally}, we release a deterministic T1--T5 taxonomy
operationalization (rule dictionaries, thresholds, and notebooks), the
field-coding rubric, and the 750-project dataset, so that future studies
can reproduce, modify, and externally validate the classification
scheme on new dormant-revived samples.
\noindent In what follows, we situate the work against prior OSS
sustainability research (§\ref{sec:related}), describe the corpus and
classifier (§\ref{sec:methodology}), report results by RQ
(§\ref{sec:results}), discuss implications (§\ref{sec:discussion}),
enumerate threats (§\ref{sec:threats}), and close with directions for
prospective validation (§\ref{sec:conclusion}).

%% file: ESEM_version/v67/3_ESEM_Edits_RelatedWork_v67.tex
\ifLIPIcsLayout\vspace{-6pt}\fi
\section{Related Work}
\label{sec:related}
\ifLIPIcsLayout\vspace{-6pt}\fi
\subsection{OSS Sustainability and Contributor Dynamics}
\ifLIPIcsLayout\vspace{-6pt}\fi
A substantial body of work has examined OSS project sustainability
through activity signals and survival. Kalliamvakou et al.\ established
that raw repository-level metrics on GitHub are unreliable proxies for
project health and can systematically misclassify both abandoned and
active projects~\cite{kalliamvakou2014promises}. Coelho and Valente
identify causes of GitHub project abandonment, operationalizing
failure as one year without commits~\cite{coelho2018modern};
Avelino et al.\ analyze 1{,}932 popular GitHub projects, finding that
a substantial fraction become abandoned within a few years but that a
non-trivial share continue under new core developers~\cite{avelino2019abandonment}.
Calefato et al.\ challenge the terminal-state assumption, showing OSS
core developers re-engage after extended absences at non-trivial
rates~\cite{calefato2022inactivity}; Iaffaldano et al.\ qualitatively
distinguish ``sleeping'' (transient) from ``dead'' (permanent)
disengagement and note the two are routinely conflated by
activity-based metrics~\cite{iaffaldano2019breaks}. Both operate at
the developer level; our study operates at the \emph{repository}
level. Miller et al.\ study how communities respond to dependency
abandonment in npm, operationalizing abandonment via a strict
two-year no-activity threshold~\cite{miller2025dependency}.
Scientific software presents a qualitatively different sustainability
problem. Howison and Herbsleb examine the incentive structures shaping
scientific software production and show that publication-driven timelines conflict with maintenance
obligations~\cite{howison2011scientific}; Howison et al.\ propose
ecosystem-level metrics arguing that individual repository signals are
insufficient~\cite{howison2015ecosystem}; Johanson and Hasselbring
characterize engineering challenges specific to computational
science~\cite{johanson2018se4science}. On contributor dynamics, Zhou
and Mockus identify factors associated with long-term OSS
participation~\cite{zhou2012longterm,zhou2015stay}; Milewicz et al.\
document high bus-factor risk in domain-specific scientific
software~\cite{milewicz2019roles}; Steinmacher et al.\ find that
documentation quality and maintainer responsiveness are critical
determinants of newcomer persistence~\cite{steinmacher2015social}.
\ifLIPIcsLayout\vspace{-12pt}\fi
\subsection{Scientific OSS Ecosystems and Post-Dormancy Trajectories}
\ifLIPIcsLayout\vspace{-6pt}\fi
Sun et al.\ conduct a mixed-methods case study of the Astropy
ecosystem~\cite{sun2025astropy}; their single-ecosystem depth
complements our cross-domain breadth. The most direct empirical
predecessor is Malviya Thakur et al., who introduce \textsc{SciCat}, a
curated corpus of 18{,}247 scientific OSS repositories drawn from World
of Code, and apply survival analysis to model project
longevity~\cite{malviyathakur2025scicat}. Using a six-month inactivity
threshold, they classify 11{,}631 projects as abandoned and note the
framework cannot distinguish abandonment from dormancy.
\ifLIPIcsLayout\vspace{-6pt}\fi
\subsection{Automated Contributions, Taxonomy Methods, and Research Gap}
\ifLIPIcsLayout\vspace{-6pt}\fi
Mirhosseini and Parnin show that dependency-update bots increase PR
merge rates but bot-generated activity does not correlate with sustained
human re-engagement~\cite{mirhosseini2017bots}; Wessel et al.\ find bots
alter perceived activity levels in ways that can mislead health
assessments~\cite{wessel2018bots}; Golzadeh et al.\ develop classifiers
to distinguish bot- from human-generated commits~\cite{golzadeh2021bots},
motivating our T4-3 (Bot-Only Pseudo-Recovery) operationalization. On
method, taxonomy-driven empirical studies are established for phenomena
that resist a single ordinal or nominal
variable~\cite{catolino2019mobilesmells,stol2018empirical,ralph2019taxonomy};
we adopt grounded
theory~\cite{strauss1990grounded,saldana2021coding,seaman1999qualitative}
for the inductive phase and the rule-based content-analysis tradition of
Krippendorff~\cite{krippendorff2004content},
Stemler~\cite{stemler2001content}, and DeFranco and
Laplante~\cite{defranco2017content} for the deductive phase.

\textbf{Research gap.}
Prior work characterizes terminal abandonment, structural fragilities
of scientific software, developer-level return patterns, and
bot-driven signal inflation, but no framework treats scientific-OSS
abandonment as a potentially recoverable state with a measurable
post-revival trajectory and explicit false-revival accounting. To our
knowledge, this paper is the first systematic study to address all
five dimensions simultaneously, grounded in 750 coded repositories
with false-revival accounting. Next, we describe the corpus,
coding protocol, and rule-based classifier that
operationalize this five-dimensional view (§\ref{sec:methodology}).

%% file: ESEM_version/v67/4_ESEM_Edits_Methodology_v67.tex
\ifLIPIcsLayout
\begin{figure}[!htbp]
\centering
\begin{minipage}[t]{0.52\linewidth}
\vspace{0pt}
\centering
\includegraphics[width=\linewidth]{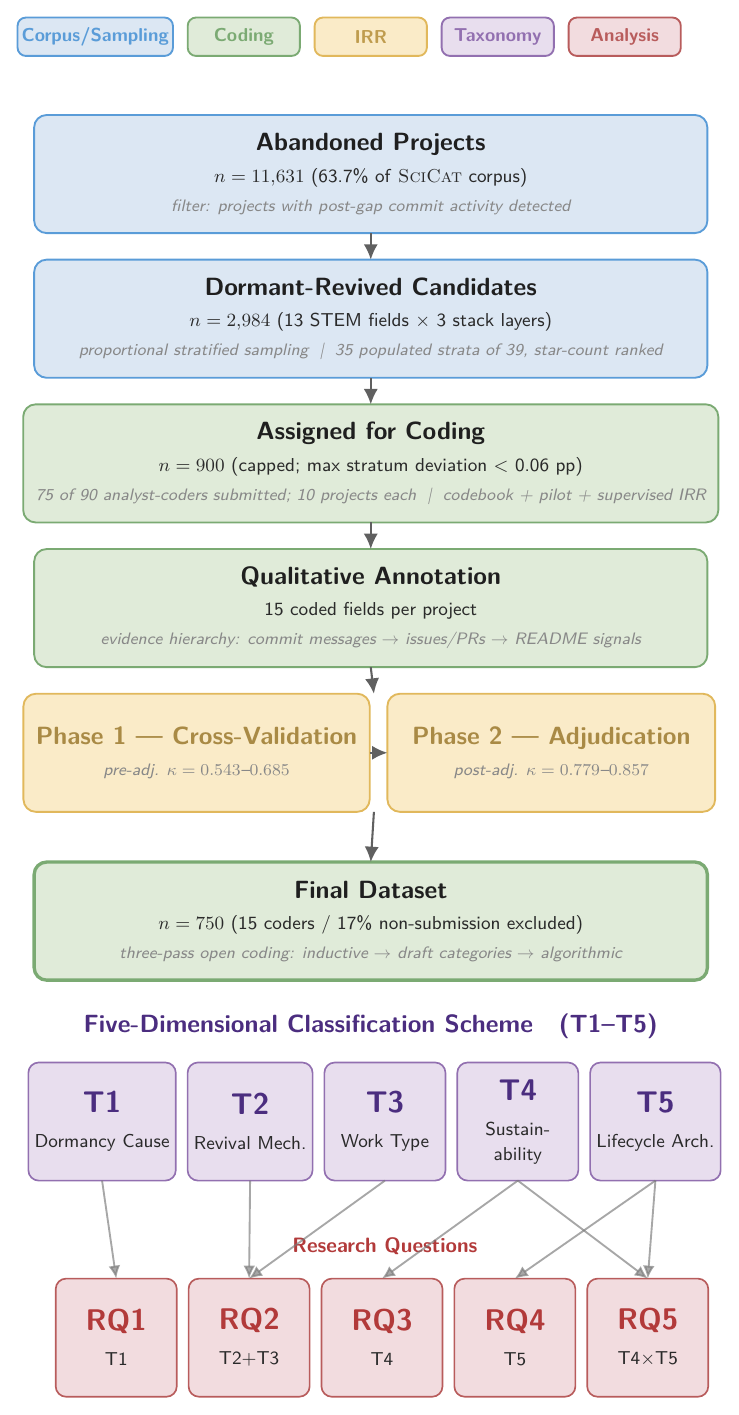}
\captionof{figure}{Method workflow for the taxonomy of dormancy and revival in scientific open-source software.
Boxes denote pipeline stages; arrows denote data flow.
}
\label{fig:methodology}
\end{minipage}\hfill
\begin{minipage}[t]{0.45\linewidth}
\vspace{-0.3in}
\centering
\scriptsize
\setlength{\tabcolsep}{2pt}
\renewcommand{\arraystretch}{0.92}
\begin{tabular}{@{}p{0.38\linewidth}p{0.57\linewidth}@{}}
\toprule
\textbf{Field} & \textbf{Role} \\
\midrule
\rowcolor{rowshade}
BeforeThemes
  & Dominant commit themes in 10 commits before the gap \\
\rowcolor{rowshade}
AfterThemes
  & Dominant themes in 10 commits after the gap ($\rightarrow$ T3) \\
\rowcolor{rowshade}
Hypothesized\-Gap\-Reason
  & Evidence-anchored interpretation of dormancy cause ($\rightarrow$ T1) \\
\rowcolor{rowshade}
HasRecovered
  & Binary recovery signal from commits, issues, or README \\
\rowcolor{rowshade}
WhyRecovered
  & Inferred revival trigger with cited repository artefacts ($\rightarrow$ T2) \\
\rowcolor{rowshade}
WhoRecovered
  & Contributor type during revival: original, new, mixed, bot ($\rightarrow$ T2, T5) \\
\rowcolor{rowshade}
CurrentStatus
  & \texttt{Active} if any commit Oct~1,~2024 -- Apr~1,~2025 \\
\bottomrule
\end{tabular}
\captionof{table}{Primary coded fields feeding T1--T5
(§\ref{sec:coding}). Numeric trajectory fields driving T4 and T5-2
thresholds (LongestGapLength, NumCommitsAfterLastGap,
NumberOfGapsInTimeline) are defined in §\ref{sec:coding}.}
\label{tab:fields}

\vspace{4pt}

\scriptsize
\setlength{\tabcolsep}{3pt}
\renewcommand{\arraystretch}{0.95}
\begin{tabular}{@{}lcccc@{}}
\toprule
\textbf{Field} & \textbf{\% Ag.} & \textbf{\% Ag.} & \boldmath$\kappa$ & \boldmath$\kappa$ \\
               & \textbf{(pre)}  & \textbf{(post)} & \textbf{(pre)}    & \textbf{(post)}   \\
\midrule
HasRecovered    & 74.9 & 88.0 & 0.543 & 0.857 \\
ActivityPattern & 72.6 & 89.3 & 0.685 & 0.820 \\
CurrentStatus   & 70.7 & 87.7 & 0.579 & 0.779 \\
RecentThemes    & 64.9 & 83.1 & N/A   & N/A   \\
WhyRecovered    & 65.7 & 85.4 & N/A   & N/A   \\
\bottomrule
\end{tabular}
\captionof{table}{Inter-rater reliability before and after adjudication for the five IRR-instrumented fields (§\ref{sec:irr}).}
\label{tab:irr}

\vspace{4pt}

\scriptsize
\setlength{\tabcolsep}{3pt}
\renewcommand{\arraystretch}{0.95}
\begin{tabular}{@{}lccc@{}}
\toprule
\textbf{Dim.} & \textbf{Raw ag.} & \boldmath$\kappa$ & \textbf{Interp.~\cite{landis1977kappa}} \\
\midrule
T2 Revival Mech.  & 74.3\% & 0.683 & Substantial \\
T4 Revival Sust.  & 76.7\% & 0.701 & Substantial \\
T5 Lifecycle Arch.& 86.6\% & 0.841 & Almost perfect \\
\bottomrule
\end{tabular}
\captionof{table}{Propagation reliability: classifier $\kappa$ on
$n=417$ IRR-paired projects. T1/T3 fields were not in formal IRR;
auxiliary check in §\ref{sec:t-label-kappa}.}
\label{tab:derived-kappa}
\end{minipage}
\end{figure}
\fi

\section{Method}\label{sec:methodology}
\ifLIPIcsLayout\vspace{-6pt}\fi
\subsection{Study Design Overview}
\ifLIPIcsLayout\vspace{-6pt}\fi
Fig.~\ref{fig:methodology} summarizes the workflow: candidate
identification from \textsc{SciCat}, stratified sampling, structured
field-level coding, two-phase inter-rater validation, and rule-based
classification into the five categorical dimensions.
The study has four critical dates: (1)~the \textsc{SciCat} corpus was
constructed and the abandonment classification applied per its published
protocol (data collection ${\sim}$July~2023); (2)~the dormant-revived
candidate set was extracted from \textsc{SciCat}'s abandoned projects
(sampling snapshot ${\sim}$August~2023, §\ref{sec:dataset}), with commit
observations bounded at October~1, 2024; (3)~qualitative coding ran
during the intervening academic semester; (4)~current-status assessment
was performed on April~1, 2025 (a project is \texttt{Active} if any
commit occurred between October~1, 2024 and April~1, 2025; six-month
window chosen to align with the \textsc{SciCat} threshold). All
findings reflect the state of the dormant-revived corpus as of
April~1, 2025. Throughout, we use \emph{sampling snapshot} for the
August~2023 re-application of the \textsc{SciCat} abandonment rule,
\emph{observation freeze} for the October~1, 2024 commit-history endpoint,
and \emph{current-status assessment} for the April~1, 2025 coder pass.

This study is a direct empirical continuation of
\textsc{SciCat}~\cite{malviyathakur2025scicat}, which classified 11,631
of 18,247 scientific OSS repositories as abandoned under a six-month
threshold while acknowledging that the framework could not distinguish
permanent abandonment from temporary dormancy. We inherit the
\textsc{SciCat} scientific-OSS definition (scientific-domain field
membership, software-stack layer, and supplementary signals) and focus
on the subset of abandoned projects that later showed renewed commit
activity (\emph{dormant-revived} or \emph{zombie} projects). The field-level coding was performed by 75 graduate and advanced
undergraduate student analyst-coders trained through a codebook, pilot
round, peer validation, and supervised adjudication, working against
a structured field-coding rubric drawn from a research-methods course
of 90 enrolled students (15 did not submit, addressed in
§\ref{sec:threats}); each coder was assigned 10 projects, yielding 750
valid primary codings. By design, each project received one primary
coding with two-phase peer validation (§\ref{sec:irr}), trading
per-project replication for broader sample coverage and structured
adjudication. T1--T5 taxonomy labels
are subsequently derived by a deterministic rule-based content-analysis
classifier (§\ref{sec:taxonomy_construction}). The remainder of
§\ref{sec:methodology} details dataset construction, the coding
protocol, inter-rater reliability, the final dataset, and the
rule-based classifier.
\subsection{Dataset Construction}
\label{sec:dataset}
\ifLIPIcsLayout\vspace{-6pt}\fi
\subsubsection{Identifying Dormant-Revived Candidates}
\ifLIPIcsLayout\vspace{-6pt}\fi
We identified candidate dormant-revived projects from \textsc{SciCat}'s
abandoned subset by re-applying its 6-month-inactivity rule at our
sampling-time snapshot ($\sim$August 2023, 7--8 weeks after the
\textsc{SciCat} paper's data-collection cutoff). At that snapshot
12,570 projects met the abandonment criterion: 939 more than the
11,631 reported in the original \textsc{SciCat} paper, due solely to
the slightly later snapshot date. Among these abandoned projects, we
flagged those that demonstrated subsequent commit activity after their
recorded last-active date, treating their abandonment label as provisional.
This yielded 2,984 candidate dormant-revived projects
spanning the scientific domains and software stack layers present in
\textsc{SciCat}. A robustness check restricting analysis to the 674
coded projects whose last commits predate the \textsc{SciCat}-paper
snapshot (n=674 of 750) yields T1--T5 distributions that match the
full-sample headline values within $\pm$1 percentage point on every
category and preserves the modal ordering on every dimension;
see §\ref{sec:threats}.

\ifLIPIcsLayout\vspace{-6pt}\fi
\subsubsection{Stratified Sampling and Coder Assignment}
\ifLIPIcsLayout\vspace{-6pt}\fi
We applied proportional stratified sampling using the cross-product of
\emph{LayerName} (Publication-Specific Code, Scientific Domain-Specific
Code, Scientific Infrastructure) and \emph{Field} (13 STEM fields),
yielding 35 populated strata of 39 possible. Within each stratum,
projects were ranked by GitHub star count, since higher-visibility
projects carry richer evidence trails (issues, PRs, README histories).
The final allocation was capped at 900 projects (maximum
stratum-proportion deviation between pool and sample $<$0.06\,pp). The
900 sampled projects were distributed at 10 projects per coder via
randomized stratum-spanning assignment so no coded set was
concentrated in a single domain or stack layer; assignments were
independent of coder demographics.

\ifLIPIcsLayout\vspace{-6pt}\fi
\subsection{Qualitative Coding Protocol and Codebook}
\label{sec:coding}
\ifLIPIcsLayout\vspace{-6pt}\fi
Our qualitative coding protocol follows the ACM SIGSOFT Empirical
Standards for Qualitative Methods~\cite{ralph2021empirical}: an
\textit{a~priori} codebook, paired-coder review of every analytic
field, structured adjudication, and pre/post-adjudication agreement
statistics reported alongside one another (§\ref{sec:irr}).

\ifLIPIcsLayout\vspace{-10pt}\fi
\subsubsection{Coding Schema}
\ifLIPIcsLayout\vspace{-6pt}\fi
Each project was coded against 15 fields (full definitions in the
replication package\footnote{\url{https://anonymous.4open.science/r/ESEM2026ReplicationPackage-598E}}).
Coders first identified the longest continuous inactivity gap in the
commit timeline, then classified each project by ecosystem layer and
field. Table~\ref{tab:fields} lists the seven primary fields feeding
T1--T5; eight supporting fields record commit-trajectory metadata and
analyst observations, of which the numeric gap-length, post-gap
commit count, and gap count drive T4 and T5-2 thresholds. The
BeforeThemes/AfterThemes/RecentThemes fields use a fixed 8-category
vocabulary (Feature Dev., Bug Fix, Docs, Dependency, Release,
Security, Bot, Other).

\ifLIPIcsLayout\vspace{-10pt}\fi
\subsubsection{Evidence Hierarchy}
\ifLIPIcsLayout\vspace{-6pt}\fi
Coders consulted evidence in priority order: (1)~commit messages (thematic and
temporal evidence); (2)~GitHub issues and PRs (dormancy context, recovery
triggers); (3)~README updates (explicit archival or revival signals). Interpretive
codings (\textit{HypothesizedGapReason}, \textit{WhyRecovered}) required citation
of specific repository artefacts (commit SHAs, PR titles).

\ifLIPIcsLayout\vspace{-6pt}\fi
\subsubsection{Analyst Training and Mitigations}
\ifLIPIcsLayout\vspace{-6pt}\fi
The 75 analyst-coders were graduate and advanced-undergraduate
students in a research-methods course supervised by a co-author. The
many-coders$\times$few-projects design (75$\times$10 = 750 codings)
maximises corpus coverage within one semester; limited per-coder
calibration is the trade-off, mitigated by: (i)~a codebook with field
definitions, decision rules, worked examples, and
\textit{ActivityPattern} reference diagrams; (ii)~a pilot-coding
revision cycle; (iii)~the two-phase IRR protocol with supervised
adjudication (§\ref{sec:irr}); and (iv)~a single supervising co-author
across all sessions. Post-adjudication agreement
($\kappa=0.779$--$0.857$, Table~\ref{tab:irr}) is substantial to almost-perfect~\cite{landis1977kappa} and exceeds the parent \textsc{SciCat} IRR ($\kappa=0.624$--$0.745$).

\ifLIPIcsLayout\vspace{-10pt}\fi
\subsubsection{Research Ethics and Student Participation}
\label{sec:ethics}
\ifLIPIcsLayout\vspace{-6pt}\fi
\textbf{Repositories, not students, are the unit of analysis.}
The codings analyzed here were produced as a graded data-coding
exercise in a graduate research-methods course; the decision to use
the dataset for research was made after the course concluded. Every
coded field is an evidence-anchored observation about a software
project, and no analytic claim concerns students as a population.
The coded dataset contains no personally-revealing coder information;
based on the U.S.\ federal Common Rule (45~CFR~46.102(e)) and our
institution's engagement-in-research criteria, we determined that this
secondary analysis does not meet the human-subjects-research
definition. Four de-identification protections apply: no
coder-identifying information in analyzed artefacts; stable arbitrary
IDs in the replication package; aggregate-only agreement reporting;
no identifiability by demographic or institutional metadata.
Following~\cite{liebel2022ethical}, we disclose that students were not
informed of potential research reuse, the exercise was graded, and no
non-coding alternative was offered.
\textit{Ethics statement for future replications.} Because research
reuse was decided only after the course concluded, no prospective IRB
review was obtained; future replications should obtain prior ethics
review, research-reuse consent, and a non-participation alternative.
Analyst-coders will be acknowledged at camera-ready.
\ifLIPIcsLayout\vspace{-10pt}\fi
\subsubsection{Imputation of Missing Fields}
\label{sec:imputation}
\ifLIPIcsLayout\vspace{-6pt}\fi
\texttt{ActivityPattern} and \texttt{CurrentStatus} were left blank by student
coders in 96 and 169 cases, respectively; both were imputed by the primary
researcher applying the identical rubric and underlying commit-frequency time
series. Post-imputation coverage: 748/750 (\texttt{ActivityPattern}) and 749/750
(\texttt{CurrentStatus}); the two remaining projects lack recoverable commit
history and are excluded from analyses requiring these fields.

\ifLIPIcsLayout\vspace{-10pt}\fi
\subsection{Inter-Rater Reliability and Adjudication}
\label{sec:irr}
\ifLIPIcsLayout\vspace{-6pt}\fi
We implemented a two-phase IRR procedure following established
practice~\cite{saldana2021coding,landis1977kappa}. In \textbf{Phase~1},
each coder independently reviewed a peer's 10 coded projects,
recording their own judgments for five key fields (\textit{HasRecovered},
\textit{ActivityPattern}, \textit{CurrentStatus}, \textit{RecentThemes},
\textit{WhyRecovered}) without access to the primary values. IRR
coverage reached 69 of 75 coder-validator pairs (92\%) and 690 of 750
projects (92\%). Assignment was randomized within the same course
section but the protocol ensured no validator had previously coded the
project they were validating, preserving project-level independence.
In \textbf{Phase~2}, pairs met in a supervised session to adjudicate
disagreements; each resolution required citation of repository
evidence and was recorded as \texttt{AuthorKept},
\texttt{ValidatorKept}, or \texttt{Compromise}.

Table~\ref{tab:irr} reports both pre- and post-adjudication agreement.
\textbf{Pre-adjudication} $\kappa$ ranged from 0.543 to 0.685 (moderate
to substantial~\cite{landis1977kappa}), with 64.9--74.9\% raw
agreement. \textbf{Post-adjudication} $\kappa$ reached 0.779--0.857
(substantial to almost perfect), with 83.1--89.3\% raw agreement,
exceeding the parent \textsc{SciCat} IRR ($\kappa=0.624$--$0.745$,
itself adjudicated) at the apples-to-apples post-adjudication stage;
pre- and post-adjudication numbers are reported separately because
our 75-coder pool is more heterogeneous than the small \textsc{SciCat}
reference team. $\kappa$ is reported only for categorical fields;
\textit{RecentThemes} and \textit{WhyRecovered} are free-text and
admit multiple valid formulations.
\ifLIPIcsLayout\vspace{-10pt}\fi
\subsubsection{Propagation Reliability of the Classifier}
\label{sec:t-label-kappa}
\ifLIPIcsLayout\vspace{-6pt}\fi
The classifier is deterministic (an assertion block verifies
reproduction of the published category counts), so variance in the
T-labels arises entirely from variance in the field inputs. To bound
that variance, we ran the classifier independently on each coder's
pre-adjudication field codings for the $n=417$ IRR-paired projects
and computed Cohen's $\kappa$ between the two derived T-label series
(Table~\ref{tab:derived-kappa}). T1 and T3 cannot be assessed this
way because their primary inputs (\textit{HypothesizedGapReason} for
T1, \textit{AfterThemes} for T3) were not in the formal IRR; an
auxiliary human-applied check covers them below.
These propagation $\kappa$ values are lower bounds because adjudication
and supervised review further reduced raw disagreement in the
production corpus. T2 and T4 depend on a small number of fields and
track the substantial-agreement band of their inputs; T5 combines
deterministic numeric signals (gap length, gap count, contributor
type) so multi-input rules average out single-field disagreement and
yield almost-perfect agreement. Sensitivity of the rule choices
themselves is reported in §\ref{sec:threats}; full confusion matrices
are in the replication-package.

\textbf{Human-applied taxonomy reliability for T1 and T3.}
To give T1 and T3 the human-applied evidence the propagation route
cannot reach, and to address concern that the taxonomy was assigned
solely by a deterministic classifier authored by a single researcher,
two researchers (the primary author and a co-author) independently
applied the T1 and T3 categories to a stratified $n=30$ sample drawn
from the 750-project corpus and balanced across stack layer and star
band. Both worked from the same field evidence the classifier reads
and were blind to one another and to the classifier's labels. Cohen's
$\kappa=0.768$ (95\% bootstrap CI $[0.59, 0.92]$, substantial per
the Landis--Koch interpretation~\cite{landis1977kappa}) on T1, with
80.0\% raw agreement; $\kappa=0.817$ (95\% CI $[0.64, 0.95]$, almost
perfect) on T3, with 86.7\% raw agreement. Coder--classifier $\kappa$
sits in the substantial band on T1 (Coder A 0.616, B 0.768) and the
moderate band on T3 (A 0.637, B 0.550); the classifier tracks human
judgment within the spread humans themselves exhibit, and the T3
moderate-band alignment bounds the strength of T3-based inferences in
§\ref{sec:rq2}. The $n=30$ sample was sized to detect $\kappa \geq 0.6$
against a null of $\kappa = 0.4$ with approximately 80\% power at
$\alpha=0.05$; bootstrap CIs are wide as expected at this $n$. Both
coders' assignments, confusion matrices, and the adjudication log are
released as \texttt{t1\_t3\_reliability.ipynb}.

\ifLIPIcsLayout\vspace{-6pt}\fi
\subsection{Final Dataset}
\label{sec:final_dataset}
\ifLIPIcsLayout\vspace{-6pt}\fi
Of the 900 projects assigned for coding, 750 valid codings were
returned. The final dataset, codebook, IRR logs, and analysis scripts
are listed in the Data Availability section.

\ifLIPIcsLayout\vspace{-6pt}\fi
\subsection{Taxonomy Construction: Rule-Based Content-Analysis Classifier}
\label{sec:taxonomy_construction}
\ifLIPIcsLayout\vspace{-6pt}\fi
The five-dimensional taxonomy (T1--T5) is the output of a deterministic
rule-based classifier applied to the coded dataset.
Table~\ref{tab:taxonomy-merged} presents the full codebook for all five
dimensions with concise category definitions; representative
repositories for all categories are in the replication-package codebook.
The classifier
(its keyword dictionaries, numeric thresholds, decision orderings, and
category-resolution rules) is fully specified in the replication
package (notebook \texttt{taxonomy\_build.ipynb}) and runs in under
two seconds on the 750-project dataset, reproducing the published
category counts exactly via a built-in assertion block.
Methodologically, this is \emph{deductive content
analysis}~\cite{krippendorff2004content,stemler2001content} applied to
software engineering by DeFranco and
Laplante~\cite{defranco2017content}, combined with inductive open
coding~\cite{strauss1990grounded} for category derivation. T-label
reliability therefore decomposes into inter-coder reliability of the
input fields (§\ref{sec:irr}; post-adjudication $\kappa =
0.779$--$0.857$) and propagation reliability of the classifier on
those inputs (§\ref{sec:t-label-kappa}). Rule-choice sensitivity is
reported in §\ref{sec:threats}.

\ifLIPIcsLayout\vspace{-6pt}\fi
\subsubsection{Inductive Phase: Rule Derivation}
\ifLIPIcsLayout\vspace{-6pt}\fi
The rule system was derived through three iterative passes: inductive
reading of all 750 \textit{Notes} and \textit{HypothesizedGapReason}
entries to surface emergent categories; collapsing them with
representative examples and operational definitions; and translating
signatures into keyword dictionaries and numeric thresholds, with
disambiguation rules codified in the notebook. \textit{Notes}
provided critical disambiguation unavailable in
\textit{HypothesizedGapReason} alone; the derivation, Notes-driven
reclassification counts, and the T1-A1 / T1-A2 split are documented
in the replication package.
\textbf{Operational thresholds.} T4-1 Sustained Recovery is
operationalized as $>$100 post-gap commits with continuing activity;
T4-4 Single-Spike as $\leq$5 post-gap commits then silence; T4-3
Bot-Only when all post-gap commits are bot-tagged with no human
re-engagement. The classifier surfaces two classes of false revival,
jointly 11.5\% of apparent revivals (T4-3 $n=14$, T4-4 $n=72$). T5
uses an ordered rule hierarchy: T5-3 (Classic Zombie) is evaluated
before T5-2 (Cyclical/Pulse). Threshold sensitivity, field coverage
(\texttt{WhyRecovered} 61.9\%, others 77.1--88.9\%), and 
 the
\texttt{WhoRecovered} free-text entries for 122 of 750 projects are
addressed in §\ref{sec:threats}.

\ifLIPIcsLayout
\begin{table}[!htbp]
\caption{Five-Dimensional Codebook (T1--T5): compact glosses. Residual
categories used in the figures and prose (T1-G Unresolvable $n=341$;
T2-U Unclear $n=192$ and T2-0 No Recovery / N/A Active $n=184$; T3-U
Unknown $n=169$; T4-X Ambiguous $n=16$; T5-X Ambiguous $n=59$ and
T5-9 Migrated $n=2$; and the N/A Active cohort $n=101$) are defined
in the replication-package codebook. Full definitions and
representative repositories are also in the package codebook.}
\label{tab:taxonomy-merged}
\label{tab:taxonomy-t1}\label{tab:taxonomy-t2}\label{tab:taxonomy-t3}\label{tab:taxonomy-t4}\label{tab:taxonomy-t5}
\centering
\scriptsize
\setlength{\tabcolsep}{4pt}
\renewcommand{\arraystretch}{0.92}
\begin{tabular}{@{}p{0.99\linewidth}@{}}
\toprule
\rowcolor{sectionshade}
\protect\dimtag{t1col}{T1}\enskip \textbf{Dormancy Cause} \\
\textbf{T1-A1:} Research Output Completion --- Paper/thesis end. \\
\textbf{T1-A2:} Feature / Milestone Freeze --- Stable release reached. \\
\textbf{T1-B:} Maintainer Departure / Burnout --- Original maintainer disengaged. \\
\textbf{T1-C:} Technical Blocker / Migration --- Major refactor or platform halt. \\
\textbf{T1-D:} Research / Academic / Funding Cycle --- Academic-calendar interruption. \\
\textbf{T1-E:} Repository Migration --- Project moved or reorganized. \\
\textbf{T1-F:} External / Planned Disruption --- External event coincident with pause. \\
\midrule
\rowcolor{sectionshade}
\protect\dimtag{t2col}{T2}\enskip \textbf{Revival Mechanism} \\
\textbf{T2-1:} Automated / Bot Revival --- First post-gap commit bot-generated. \\
\textbf{T2-2:} Original Maintainer Return --- Original maintainer resumed. \\
\textbf{T2-3:} New Contributor Injection --- New contributor initiated revival. \\
\textbf{T2-4:} Technical / Compatibility Trigger --- Dependency/platform forced re-engagement. \\
\textbf{T2-5:} Research / Feature Renewal --- New feature or research direction. \\
\textbf{T2-6:} Release-Driven Revival --- Release/CI preparation work. \\
\textbf{T2-7:} Maintenance / Documentation Only --- Housekeeping/docs commits only. \\
\midrule
\rowcolor{sectionshade}
\protect\dimtag{t3col}{T3}\enskip \textbf{Nature of Revival Work} \\
\textbf{T3-1:} Feature Development --- New modules or API surface. \\
\textbf{T3-2:} Bug-Fix / Corrective --- Fix-only commits. \\
\textbf{T3-3:} Dependency Modernization --- Dependency/compatibility updates. \\
\textbf{T3-4:} Mixed (Feature + Maint.) --- Features plus maintenance. \\
\textbf{T3-5:} Documentation / Housekeeping --- Docs/housekeeping only. \\
\textbf{T3-6:} Bot-Automated Only --- Exclusively bot commits. \\
\textbf{T3-7:} Release / Versioning --- Version bumps, changelog. \\
\midrule
\rowcolor{sectionshade}
\protect\dimtag{t4col}{T4}\enskip \textbf{Revival Sustainability} \\
\textbf{T4-1:} Sustained Recovery --- $>100$ post-gap commits, continuing. \\
\textbf{T4-2:} Partial / Fragile Recovery --- Modest post-gap commit volume. \\
\textbf{T4-3:} Bot-Only Pseudo-Recovery --- Only bot commits post-gap. \\
\textbf{T4-4:} Single-Spike ($\leq 5$ commits) --- $\leq 5$ commits then silence. \\
\textbf{T4-5:} Recovered-Then-Declined --- Genuine activity that faded. \\
\midrule
\rowcolor{sectionshade}
\protect\dimtag{t5col}{T5}\enskip \textbf{Lifecycle Archetype} \\
\textbf{T5-1:} Evergreen --- Continuously active. \\
\textbf{T5-1b:} Active, Short Gap --- Short gap, sustained recovery, active. \\
\textbf{T5-2:} Cyclical / Pulse --- Recurring burst-and-silence cycles. \\
\textbf{T5-3:} Classic Zombie --- Multi-year dormancy, sustained revival. \\
\textbf{T5-4:} Rescue / Handoff --- New contributor took over. \\
\textbf{T5-5:} Phantom Revival --- Brief burst then silence resumes. \\
\textbf{T5-6:} Bot Zombie --- Bot-only post-gap activity. \\
\textbf{T5-7:} Transient / Paper-Code --- Paper-companion; ends after publication. \\
\textbf{T5-8:} Declining --- Recovered then trending inactive. \\
\bottomrule
\end{tabular}
\end{table}

\fi

%% file: ESEM_version/v67/5_ESEM_Edits_Results_Section_v67.tex
\ifLIPIcsLayout\vspace{-6pt}\fi
\section{Results}
\label{sec:results}
\ifLIPIcsLayout\vspace{-6pt}\fi

\ifFigDistributions
\begin{figure*}[t]
\vspace{-.2in}
  \centering
  \includegraphics[width=0.99\textwidth]{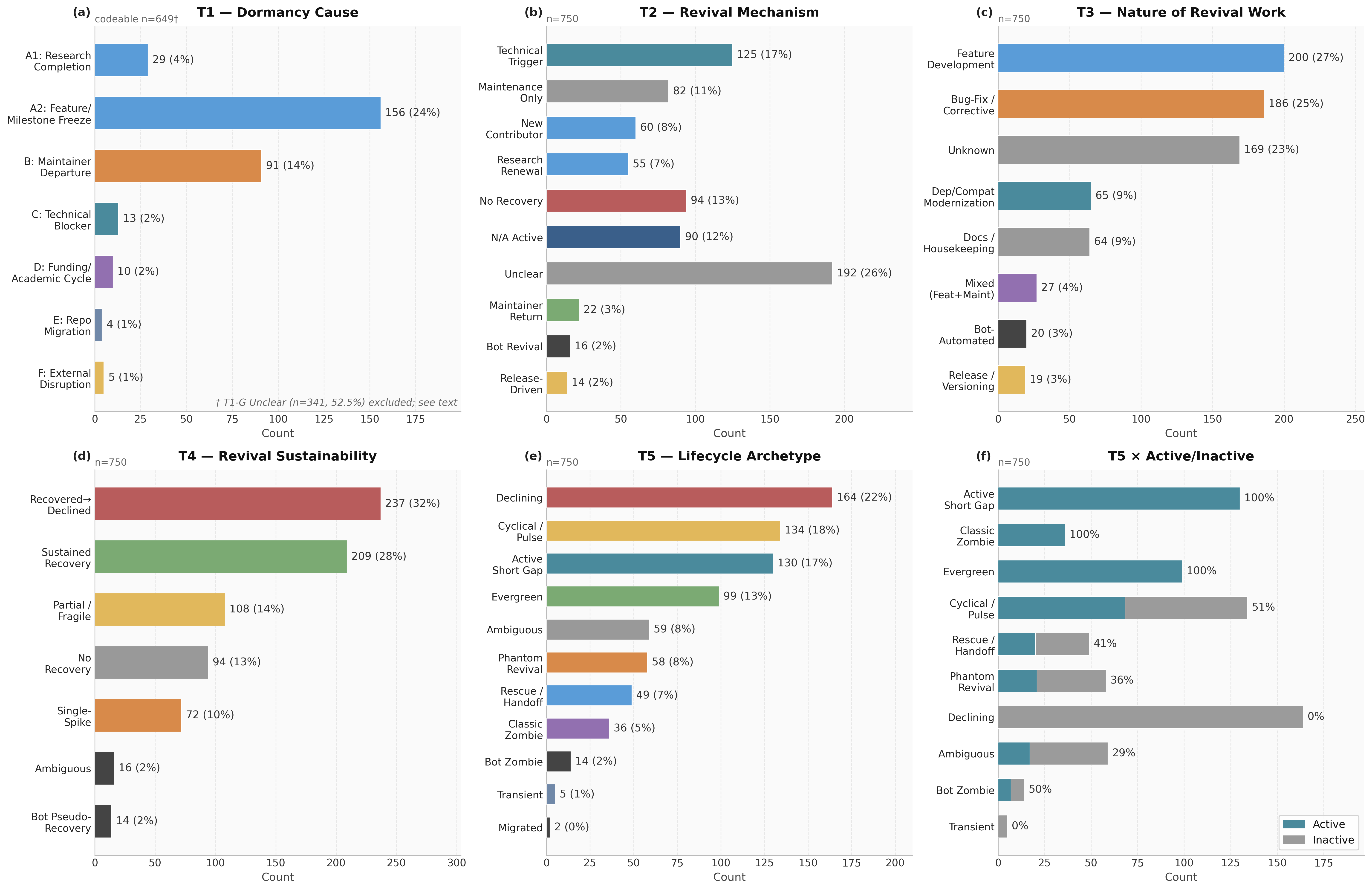}
\vspace{-.05in}
  \caption{Taxonomy category distributions ($n=750$). (a) T1 dormancy cause
  (codeable $n=649$; T1-G excluded, see text). (b) T2 revival mechanism.
  (c) T3 nature of revival work. (d) T4 revival sustainability.
  (e) T5 lifecycle archetype. (f) T5 archetype by current status.}
  \label{fig:distributions}
 \vspace{-.05in}
\end{figure*}
\fi
We present results for five research questions, each addressed by one or more
taxonomy dimensions (T1--T5)\ifFigDistributions; Fig.~\ref{fig:distributions} reports
the distribution of all 46 categories across $n=750$ projects\fi.
All reported rates are conditional on our visibility-weighted coded
sample (median 170 GitHub stars vs.\ 9 in the \textsc{SciCat} abandoned
parent population); star-band sensitivity is in
Table~\ref{tab:starband} (§\ref{sec:threats}).
Because different analyses use different subsamples (full $n=750$,
codeable $n=649$, non-imputed $n=515$, focal $5\times4$ $n=511$,
structurally-independent $n=266$), the full denominator cascade from
the \textsc{SciCat} parent corpus to each analysis subsample is
documented in the replication package, so each reported $n$ in
§\ref{sec:results} can be located against the cascade.
Three findings form a coherent
associative chain that runs through RQ3--RQ5: \emph{longer dormancy is associated
with a lower-sustainability archetype, lower-sustainability archetypes are
associated with non-sustained outcomes, and gap duration is independently
associated with outcome}. Revival
mechanism (T2) and work type (T3) show no significant association with this
chain at our available statistical power, and their direct practical implications
are discussed in §\ref{sec:discussion}.

\ifLIPIcsLayout\vspace{-6pt}\fi
\subsection{RQ1. Causes of Extended Dormancy}
\label{sec:rq1}
\ifLIPIcsLayout\vspace{-6pt}\fi

\ifFigDistributions
Fig.~\ref{fig:distributions}(a) shows the T1 distribution across the 649
codeable projects (101~N/A active projects have no dormancy gap to classify).
\fi
\textbf{Measurement-validity ceiling: cause is unresolvable for
the majority.} T1-G (Unclear / Insufficient Evidence) accounts for
341 of 649 codeable projects (52.5\%): repository-visible evidence
alone cannot support confident dormancy-cause attribution for the
majority of scientific OSS projects in our sample, reflecting the
brief commit messages and limited issue-tracker activity typical of
single-maintainer scientific repositories. We treat this 52.5\% as a
quantified bound on what automated dormancy-cause attribution from
commit-trace signals alone can achieve, not merely as a study
limitation.
\textbf{Among resolvable causes, feature freeze dominates
research completion 5.4:1.} T1-A2 (Feature / Milestone Freeze)
accounts for 156 of the 308 projects with identifiable causes
(50.6\% of resolvable; 24.0\% of all codeable), the most common
named cause by a wide margin. T1-A1 (Research Output
Completion), the theoretically expected dominant cause given the
publication-driven model of scientific
software~\cite{malviyathakur2025scicat}, ranks third at 4.5\%
($n=29$), behind T1-B (Maintainer Departure / Burnout) at 14.0\%
($n=91$). The 5.4:1 freeze-to-completion ratio is surprising: it does
not estimate population prevalence of true dormancy causes but
characterizes what the visible evidence supports when evidence is
sufficient: among cases where repository evidence
supports an interpretation, scientific OSS appears more often to stop
because it reached a deliberate checkpoint than because the underlying
research effort ended.
\textbf{Cause shows no detectable association with duration.}
Kruskal-Wallis across the seven 
T1 categories: $H=1.76$, $df=6$, $p=0.940$,
$\varepsilon^2\approx 0.000$ (negligible). Median gap range: 4.0~mo (T1-C Technical Blocker) to 9.0~mo (T1-B Maintainer Departure).
The T1 majority-unresolvable result is robust in direction but
coder-sensitive in exact magnitude: excluding 14 high-T1-G coders
shifts the unresolvable rate from 52.5\% to 42.2\% (§\ref{sec:threats}).
ceiling on dormancy-cause attribution, not as a precise population estimate.
We therefore treat T1 less as a precise population estimate and more as evidence that repository data often limits how confidently dormancy causes can be attributed.

\ifLIPIcsLayout\vspace{-6pt}\fi
\subsection{RQ2. What Mechanisms Trigger Revival, and What Work Characterizes Recovery?}
\label{sec:rq2}
\ifLIPIcsLayout\vspace{-6pt}\fi
RQ2 addresses two complementary facets of the revival event: the external
trigger \ifFigDistributions(T2, Fig.~\ref{fig:distributions}(b))\fi\
and the nature of resumed
work \ifFigDistributions(T3, Fig.~\ref{fig:distributions}(c))\fi.
\textbf{Descriptive findings.}
T2-4 (Technical / Compatibility Trigger) is the most common
identifiable revival mechanism (16.7\%, $n=125$, more than double the
next identifiable category); T3-1 (Feature Development) leads T3
(26.7\%, $n=200$), narrowly ahead of T3-2 (Bug-Fix / Corrective, 24.8\%,
$n=186$); post-gap work appears to mix renewed development with
accumulated maintenance, consistent with a combination of feature-cycle
revival and technical-debt repayment rather than either alone. T2-3 (New Contributor Injection, $n=60$, 8.0\%) outnumbers
T2-2 (Original Maintainer Return, $n=22$, 2.9\%) at 2.7:1: in the
coded revival mechanisms, new-contributor-triggered revival is more
common than explicit original-maintainer return.
Maintenance-only revivals (T2-7, 10.9\%, $n=82$) provide weaker
evidence of deep codebase recovery or renewed feature development and
may be over-counted as full recovery by na\"ive commit-presence metrics.
\textbf{Temporally ordered hypothesis test: mechanism non-significant,
work type weak and inconclusive.}
We tested the temporally ordered hypothesis that revival mechanism (T2)
and revival work type (T3) are associated with T4 sustainability at
$\alpha=0.05$. At $\alpha=0.05$, revival mechanism is non-significant
(T2$\times$T4 excluding No Recovery / Unclear rows: $\chi^2=16.05$,
$df=12$, $p=0.189$, $V=0.158$), while work type shows weak and
inconclusive evidence of association with T4
(T3$\times$T4: $\chi^2=28.81$, $df=18$, $p=0.051$, $V=0.136$). With $n=344$, this test has approximately 80\% power to
detect $V \geq 0.15$ at $\alpha=0.05$; the observed $V=0.158$ sits at
that detection floor, so a true effect, if present, is bounded above by
small magnitude. Given sparse cells and the multi-category table, we
therefore interpret the non-significant T2 result as no robust detectable
association rather than confirmed independence, and treat proximate
revival characteristics as weaker and less robust sustainability signals
than lifecycle archetype and gap duration (RQ4, RQ5). Because T3 has
only auxiliary human reliability evidence on $n=30$ (§\ref{sec:coding})
and moderate coder--classifier alignment, the T3 association result
should be interpreted as descriptive and exploratory rather than as a
validated automated work-type classifier.

\ifLIPIcsLayout\vspace{-6pt}\fi
\subsection{RQ3. How Often Does Apparent Abandonment Lead to Meaningful,
Sustained Recovery?}
\ifLIPIcsLayout\vspace{-6pt}\fi
\label{sec:rq3}
\ifFigDistributions
Fig.~\ref{fig:distributions}(d) shows the T4 distribution and
Fig.~\ref{fig:boxplots} shows gap duration by outcome.
\fi
\textbf{Apparent revival is not predominantly a story of restoration
but of temporary reprieve.} Non-sustained outcomes outnumber Sustained
Recovery 2.14:1 in the full sample and 1.74:1 in the 515 non-imputed
projects (under our visibility-weighted sampling design); aggregated
rank is stable across imputation choice. T4-5 (Recovered-Then-Declined,
31.6\%, $n=237$) and T4-1 (Sustained Recovery, 27.9\%, $n=209$) are
the two largest outcome categories; T4-1 is the most band-sensitive
headline (§\ref{sec:threats}, Table~\ref{tab:starband}).
\textbf{False revivals inflate apparent recovery rates by 11.5 percentage points.}
14 projects show bot-only activity (T4-3) and 72 show single-spike
events of five or fewer commits (T4-4). The 11.5\% rate is band-stable
across star-count strata (Table~\ref{tab:starband}); studies or tools
relying on raw post-gap commit presence will overestimate meaningful
revival rates unless they separate bot-only and single-spike activity.
\textbf{Shorter dormancy is the earliest signal associated with sustained recovery.}
Kruskal-Wallis across four T4 groups: $H=69.35$, $df=3$, $p=5.87\times10^{-15}$,
$\varepsilon^2=0.104$ (medium effect). Median gap lengths exhibit a clear
gradient: T4-1 Sustained~=~3.0~mo, T4-5~Declined~=~6.0~mo,
T4-2~Partial~=~7.5~mo, T4-0~No~Recovery~=~9.0~mo (Fig.~\ref{fig:boxplots}).
Pairwise Mann-Whitney tests confirm Sustained recovery is significantly separated
from all other outcomes ($p<0.001$, Bonferroni-corrected). Dormancy beyond
approximately six months is associated with substantially lower
observed durable-recovery rates.
\textbf{Sustained and declined recoveries differ substantially in depth.}
Median post-gap commits differ by ${\sim}7.8\times$ (T4-1 Sustained~=~1{,}604
vs.\ T4-5 Declined~=~206): sustained recovery is characterized by deep,
ongoing re-engagement rather than a short burst of catch-up commits
followed by renewed inactivity. Because post-gap commit volume is part
of the T4-1 operational definition, this depth comparison is a
descriptive check on operationalization rather than independent
validation of sustainability.
\textbf{Same-contributor revival predominates.}
Among 389 of 750 projects with codebook-compliant \texttt{WhoRecovered}
entries, same-contributor returns outnumber new-contributor handoffs
at 2.7:1; an audit of free-text entries gives 3.3:1 on $n=452$
(notebook \texttt{whorecovered\_audit.ipynb}), with 2.7:1 retained as
conservative. The 2.7:1 ratio is numerically similar to the RQ2
T2-3:T2-2 ratio but captures a distinct construct: T2 codes the
inferred trigger mechanism, \texttt{WhoRecovered} codes contributor
identity during recovery.

\ifLIPIcsLayout\vspace{-6pt}\fi
\subsection{RQ4. What Lifecycle Archetypes Characterize Dormant-Revived
Scientific OSS Projects?}
\label{sec:rq4}
\ifLIPIcsLayout\vspace{-6pt}\fi
\ifFigDistributions
Fig.~\ref{fig:distributions}(e)(f) show the T5 archetype distribution and
archetype-by-current-status breakdown.
\fi

\textbf{Declining is the dominant archetype, but healthy trajectories
remain common, and dormancy duration distinguishes archetypes with
large effect.} T5-8 Declining (21.9\%, $n=164$) is the single most
prevalent archetype, but healthy archetypes together exceed it: T5-1b
Active Short Gap (17.3\%) plus T5-2 Cyclical/Pulse (17.9\%) account
for 35.2\% of the corpus, and T5-1 Evergreen (13.2\%) adds a further
no-dormancy cohort.\footnote{T5-1 Evergreen projects entered the
dormant-revived candidate set via SciCat's snapshot-date inactivity
heuristic but, on manual inspection of the full commit timeline,
show no gap meeting our manual dormancy threshold; they are
artefacts of the SciCat snapshot date rather than true dormant
projects. Throughout the dormancy and recovery analyses they are
interpreted as no-gap reference cases rather than as dormant-revived
projects. In §\ref{sec:rq5} they are retained in the
structurally-independent V analysis specifically as a no-gap baseline
cell, since their T4 distribution is logically separate from T4 by
construction; a sensitivity rerun without them is reported there.} T5-3 Classic Zombie
(4.8\%, $n=36$, 95\% CI
[3.5\%, 6.6\%]) provides concrete counterexamples to the assumption that extended
inactivity necessarily implies abandonment; all 36 Classic Zombie projects are
currently active. Archetype labels align with observed status (T5-1b
and T5-3 = 100\% active; T5-8 = 0\% active; T5-2 = 51\% active),
consistent with the taxonomy's face validity. Kruskal-Wallis across six focal
archetypes shows large-effect dormancy-duration discrimination
($H=198.66$, $df=5$, $p=5.50\times10^{-41}$, $\varepsilon^2=0.346$;
median gap span 3.0~mo for T5-1b through 16.5~mo for T5-3); archetype
membership explains approximately one-third of gap-duration variance.

\ifLIPIcsLayout
\begin{table}[!htbp]
\centering
\begin{minipage}[t]{0.48\linewidth}
\vspace{0pt}
\centering
\scriptsize
\setlength{\tabcolsep}{2pt}
\renewcommand{\arraystretch}{0.95}
\begin{tabular}{@{}p{0.40\linewidth}rcccc@{}}
\toprule
\textbf{Archetype (T5)} & \textbf{$n$} &
  \textbf{T4-1} & \textbf{T4-2} & \textbf{T4-5} & \textbf{T4-0} \\
  & & \textbf{Sust.} & \textbf{Part.} & \textbf{Decl.} & \textbf{None} \\
\midrule
T5-1b Active Short Gap & 130 & \textbf{65.4\%} & 28.5\% &  0.0\% &  6.2\% \\
T5-2 Cyclical/Pulse    & 134 & 30.6\%          & 20.1\% & 37.3\% & 11.9\% \\
T5-3 Classic Zombie    &  36 & \textbf{63.9\%} & 30.6\% &  0.0\% &  5.6\% \\
T5-4 Rescue/Handoff    &  48 & 20.8\%          & 16.7\% & \textbf{58.3\%} &  4.2\% \\
T5-8 Declining         & 163 &  0.0\%          &  5.5\% & \textbf{74.8\%} & 19.6\% \\
\bottomrule
\end{tabular}
\captionof{table}{T5 lifecycle archetype $\times$ T4 sustainability
outcome (row percentages; $n=511$, focal $5\times4$ sub-table).
$\chi^2=277.72$, $df=12$, $p<0.001$, $V=0.426$. Restricted to focal
T5$\times$T4; row counts may differ from Fig.~\ref{fig:distributions}(e)
by $\pm1$.}
\vspace{-2pt}
\label{tab:rq5_crosstab}
\end{minipage}\hfill
\begin{minipage}[t]{0.45\linewidth}
\vspace{0pt}
\centering
\scriptsize
\setlength{\tabcolsep}{2pt}
\renewcommand{\arraystretch}{0.99}
\begin{tabular}{lrrrrrr}
\toprule
\textbf{Band} & \boldmath$n$ & \boldmath$n_{\text{cod.}}$ & \textbf{T1-G} & \textbf{T4-1} & \textbf{T4-5} & \textbf{False} \\
 & & & \textbf{unr.} & \textbf{Sust.} & \textbf{Decl.} & \textbf{rev.} \\
\midrule
Low ($<$50)     & 196 & 180 & 51.1\% & 16.3\% & 35.2\% & 15.3\% \\
Mid (50--499)   & 309 & 272 & 52.2\% & 32.4\% & 27.2\% &  9.7\% \\
High ($\geq$500)& 219 & 179 & 53.6\% & 29.7\% & 33.8\% & 11.9\% \\
\midrule
All             & 750 & 649 & 52.5\% & 27.9\% & 31.6\% & 11.5\% \\
\bottomrule
\end{tabular}
\captionof{table}{Headline rates stratified by GitHub star count band
within the coded corpus ($n=750$; T1-G computed on $n=649$ codeable
projects). False-revival rate is T4-3 plus T4-4.}
\vspace{-2pt}
\label{tab:starband}
\vspace{-2pt}
\end{minipage}
\vspace{-2pt}
\end{table}
\ifLIPIcsLayout\vspace{-6pt}\fi
\begin{figure}[!htbp]
\centering
\begin{minipage}[t]{0.48\linewidth}
\vspace{0pt}
\centering
\includegraphics[width=\linewidth]{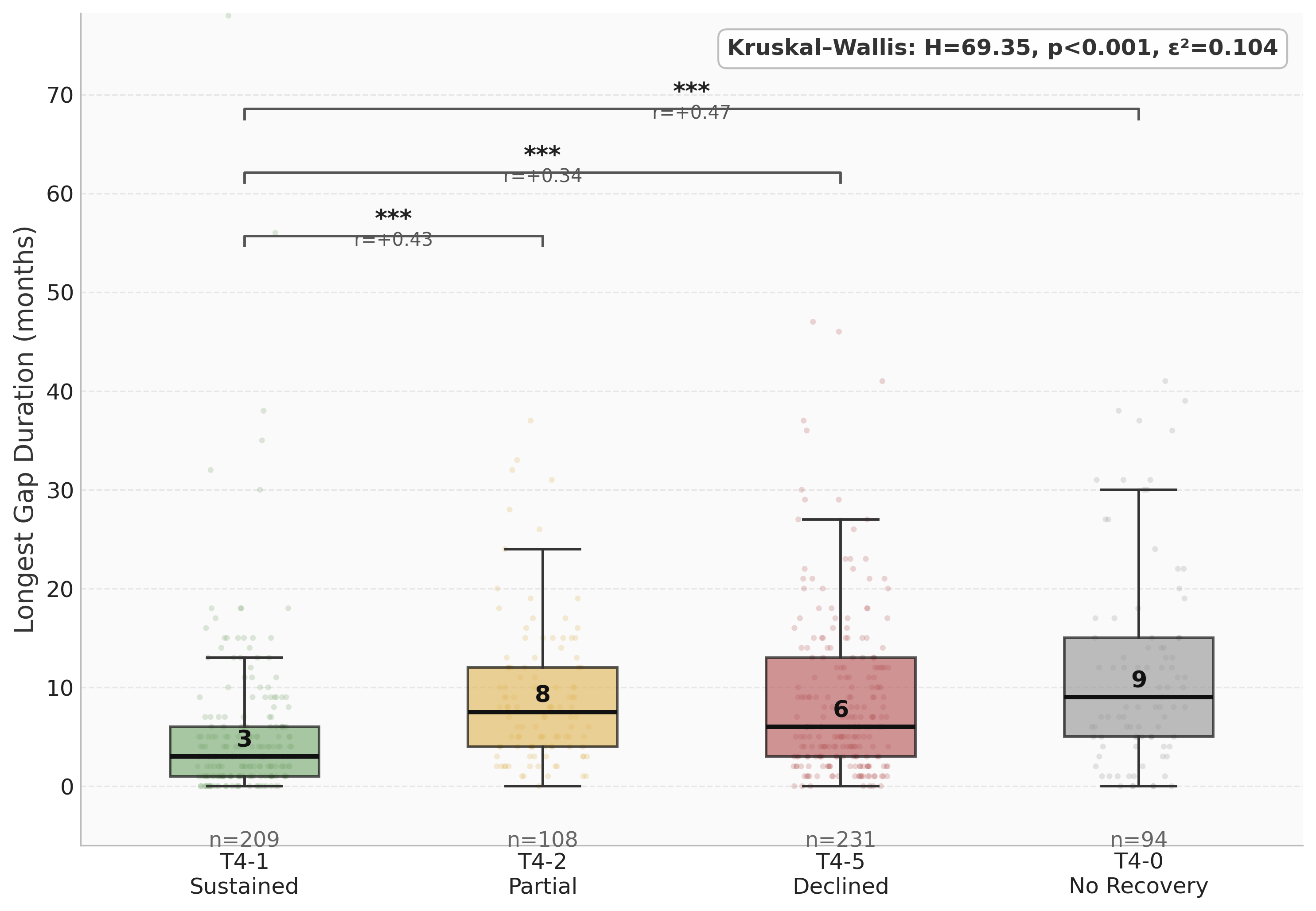}
\captionof{figure}{Gap duration by revival sustainability outcome.
Median values labelled inside boxes; jittered dots show individual
projects; $r$ values under significance brackets give rank-biserial
effect sizes. Bonferroni-corrected pairwise Mann--Whitney tests
(*** $p<0.001$). Kruskal--Wallis: $H=69.35$, $p<0.001$,
$\varepsilon^2=0.104$. Excludes T4-3/T4-4/T4-X and cases lacking gap
data (vs.\ Fig.~\ref{fig:distributions}(d)).}
\label{fig:boxplots}
\end{minipage}\hfill
\begin{minipage}[t]{0.48\linewidth}
\vspace{0pt}
\centering
\includegraphics[width=\linewidth,trim={2.55cm 0cm 3.5cm 0cm},clip]{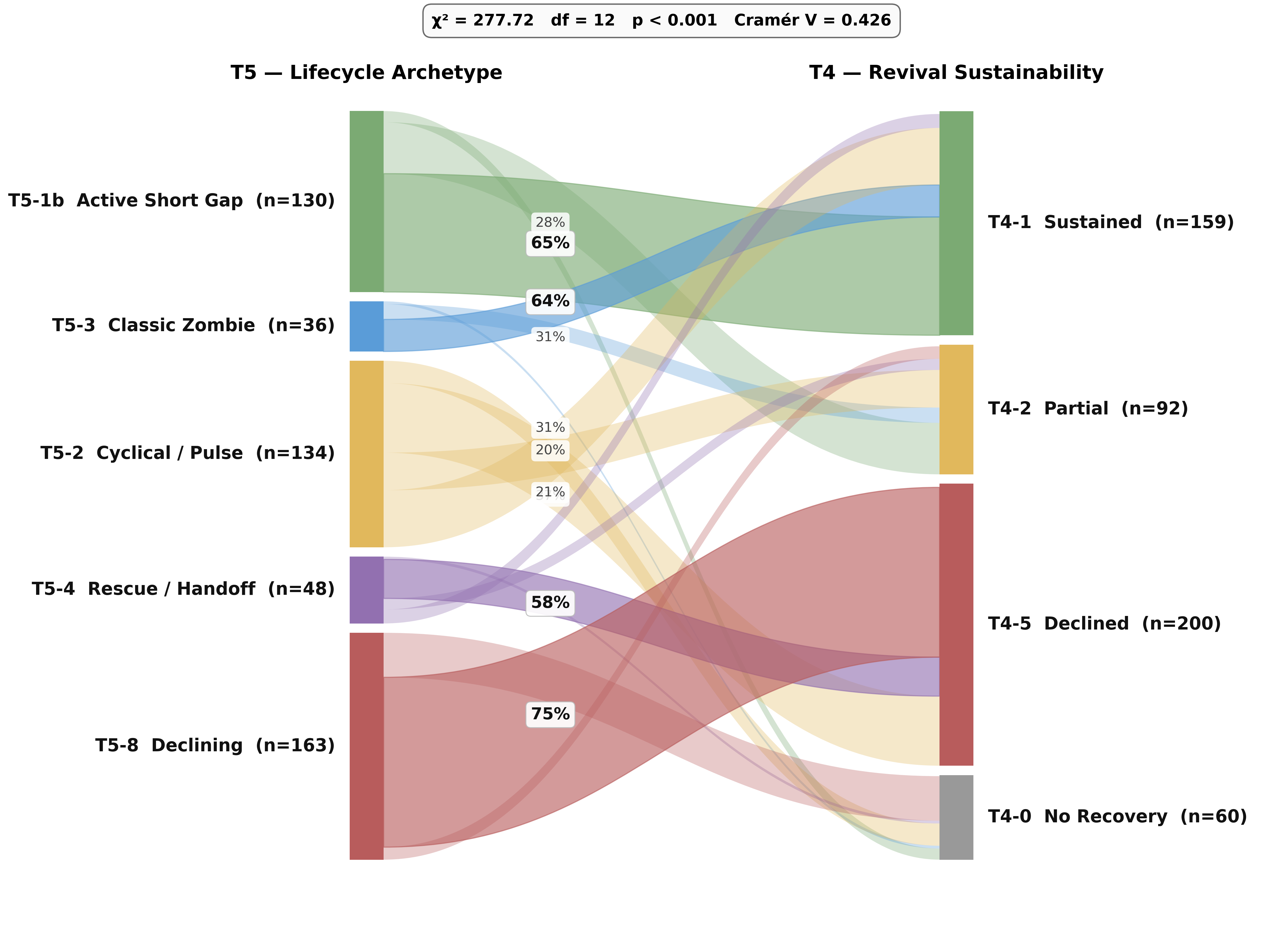}
\captionof{figure}{Lifecycle archetype (T5) to revival sustainability
outcome (T4) for the focal $5\times4$ sub-table ($n=511$). Flow width
indicates project count; bold labels mark dominant row flows
($\geq 50\%$). Row percentages and statistics in
Table~\ref{tab:rq5_crosstab}.}
\label{fig:sankey}
\end{minipage}
\end{figure}
\fi
\ifLIPIcsLayout\vspace{-12pt}\fi
\subsection{RQ5. How Do Lifecycle Archetypes Align with Revival Sustainability?}
\label{sec:rq5}
\ifLIPIcsLayout\vspace{-6pt}\fi

Fig.~\ref{fig:sankey} shows the T5$\to$T4 flow ($n=511$ after restriction
to the focal $5\times4$ sub-table: T5 $\in$ \{T5-1b, T5-2, T5-3, T5-4,
T5-8\} and T4 $\in$ \{T4-0, T4-1, T4-2, T4-5\}; structurally orthogonal
categories T5-1, T5-5, T5-6, T5-7 and T4-3, T4-4 are excluded, as are
the residual T5-9, T5-X, and T4-X  defined in the
replication-package codebook). Table~\ref{tab:rq5_crosstab} reports row percentages.
\textbf{In this operationalization, archetype shows a stronger sustainability association than the tested T2/T3 variables.}
On the structurally-independent archetype subset (T5 $\in$ \{T5-1, T5-2,
T5-4\}, $n=266$; see the construct-overlap discussion below in this
subsection), $V=0.268$ (medium effect, $p<10^{-5}$). T5-1 (Evergreen)
is retained here as a no-gap baseline cell whose T4 distribution is
logically separate from T4 by construction (its inclusion is the
treatment consistent with the no-real-dormancy footnote in
§\ref{sec:rq3}: T5-1 is interpreted as a no-gap reference, not as
dormant-revived). A sensitivity rerun excluding T5-1 (T5 $\in$ \{T5-2,
T5-4\}, $n=183$) yields $V=0.233$, $\chi^2=9.91$, $df=4$, $p=0.042$,
still a medium effect; the result is robust to the T5-1 inclusion
choice. On the full focal sub-table ($n=511$), $\chi^2=277.72$,
$df=12$, $p=2.20\times10^{-52}$, Cram\'{e}r's $V=0.426$; the
full-table $V$ is inflated by construct overlap with T4 conditions
built into some T5 categories and is best read as taxonomy-alignment
evidence rather than independent predictive strength. On the
structurally-independent subset, archetype is more strongly associated
with sustainability outcome than revival mechanism ($V=0.158$,
$p=0.189$) or work type ($V=0.136$, $p=0.051$).
\textbf{Archetype-level patterns.} Across the focal sub-table
(Table~\ref{tab:rq5_crosstab}): Active Short Gap (T5-1b) and Classic
Zombie (T5-3) both sustain at approximately two-thirds despite a
5.5$\times$ difference in median gap duration; Declining (T5-8) shows
the most consistent negative outcome (almost no sustained recoveries);
Rescue/Handoff (T5-4) declines at the highest rate of any recovery
archetype, consistent with the hypothesis that domain-knowledge transfer
is a barrier to durable revival under new contributors; and
Cyclical/Pulse (T5-2) shows a near-even split between sustained and
declined outcomes, the highest-uncertainty class.
\textbf{T4 and T5 are not fully independent constructs.} Several T5 archetypes
are defined in terms that overlap with T4 outcome conditions: T5-1b
(Active Short Gap) and T5-3 (Classic Zombie) are operationalized around
sustained recovery and so cluster heavily on T4-1 (65.4\% and 63.9\%
respectively in Table~\ref{tab:rq5_crosstab}); T5-5 = T4-4 (Phantom
Revival is the single-spike pattern); T5-6 $\subseteq$ T4-3 (Bot Zombie
is bot-only recovery); T5-7 $\subseteq$ T4-0 (Transient/Paper-Code
typically shows no post-publication recovery); T5-8 $\subseteq$ T4-5
(Declining is the inactive-after-recovery pattern). Only T5-1
(Evergreen, as a no-gap baseline), T5-2 (Cyclical/Pulse), and T5-4
(Rescue/Handoff) stand on dimensions logically separate from T4.
Restricting the test to these three structurally-independent archetypes
($n=266$, T5 $\in$ \{T5-1, T5-2, T5-4\}; T4 $\in$ \{T4-0, T4-1, T4-2,
T4-4, T4-5\}) yields $\chi^2=38.08$, $df=8$, $p<10^{-5}$, $V=0.268$
(medium effect); dropping T5-1 entirely yields $V=0.233$, $p=0.042$
($n=183$), still medium. The full-table $V=0.426$ is therefore inflated
by structural overlap, but a substantial archetype--outcome association
remains in the structurally-independent subset; the mechanical coupling
does not account for the entire effect. This is discussed further as a
construct-validity threat in §\ref{sec:threats}.
\textbf{Synthesis: an associative chain from dormancy to outcome.}
The three significant findings across RQ3--RQ5 form a coherent
\emph{associative, not causal} chain: longer dormancy is associated
with lower-sustainability archetype (gap-to-archetype:
$H=198.66$, $\varepsilon^2=0.346$), which is associated with
non-sustained outcome (archetype-to-outcome on the
structurally-independent subset: $V=0.268$, see above); gap duration
also relates independently to outcome ($H=69.35$,
$\varepsilon^2=0.104$). Revival mechanism (T2) and work type (T3) are
largely orthogonal to this chain. The observational design cannot
distinguish a causal chain, a common-cause structure, or independent
effects. A practical takeaway worth prospective validation: projects
reviving from $\leq$3~months sustain recovery at 65\% (T5-1b); those
reviving from $\geq$12~months either recover durably (T5-3: 64\%
sustained) or are already declining (T5-8: 75\% declined), with little
middle ground. In the next section we interpret these findings for
ecosystem health monitoring and position them against general-OSS
results (§\ref{sec:discussion}).

%% file: ESEM_version/v67/6_ESEM_Edits_Discussion_v67.tex
\section{Discussion}
\label{sec:discussion}
\ifLIPIcsLayout\vspace{-6pt}\fi
\subsection{What Happens After Apparent Abandonment  and Why Thresholds Fail}
\ifLIPIcsLayout\vspace{-6pt}\fi
The motivating question receives a two-sided answer within our
dormant-revived candidate frame: we find that renewed activity is common enough to
undermine binary inactivity labels, but durable recovery is not the
dominant observed trajectory. The
issue, then, is not that inactivity thresholds are too lax or too
strict, but that a single duration cut-off conflates trajectories that
observed outcomes separate cleanly. Ecosystem health assessments that
treat raw post-gap activity as recovery, without distinguishing
sustained from temporary revival, risk overstating scientific software
health. Our reported rates reflect a high-visibility sample (median
170 stars); among the more numerous low-visibility projects,  false
-revivals are likely more prevalent than the 11.5\% we observe
(§\ref{sec:threats}).
A single inactivity threshold appears insufficient to reliably classify projects as
abandoned because we observe that outcomes associate with factors a
threshold does not capture. \textbf{Gap duration matters non-linearly}:
short gaps are associated with sustained recovery, longer gaps with
decline, yet T5-3 Classic Zombie projects revive after gaps exceeding
twelve months, showing that extended inactivity is not a reliable
abandonment signal. \textbf{Lifecycle archetype} (RQ4, RQ5) provides
the discriminating signal duration cannot: knowing trajectory type
substantially narrows plausible outcomes, and a threshold-based
classifier cannot make this distinction during the dormancy window.
The multivariate robustness check (§\ref{sec:threats}) confirms the
T5 signal survives controls for scale, popularity, and post-gap commit
volume (T5-3 OR~$=8.16$, $p<10^{-3}$; $\log(\text{stars}+1)$ not
significant). \textbf{Contributor continuity} adds a third dimension:
Rescue/Handoff projects show the highest decline rate of any recovery
archetype, suggesting that scientific OSS may encode domain knowledge
not easily transferred through repository metadata alone (a hypothesis
for future practitioner validation). By contrast, revival mechanism
(T2, $p=0.189$) and resumed work type (T3, $p=0.051$, $V=0.136$) are
weaker signals than archetype and gap duration in this sample:
structural properties established \emph{before} the revival event
appear more informative than proximate revival mechanism or work type.
\ifLIPIcsLayout\vspace{-6pt}\fi
\subsection{Implications and Position vs Prior Work}
\ifLIPIcsLayout\vspace{-6pt}\fi
The taxonomy suggests design hypotheses for three stakeholder groups, offered as candidates.
\textbf{For Tool Builders and Health-Dashboard Developers.}
Tools that infer dormancy cause could expose unresolvable evidence
rather than commit to a label, given the T1-G majority (52.5\%,
§\ref{sec:rq1}). The 11.5\% false-revival rate (T4-3 + T4-4) is a
lower bound on misclassification by any commit-presence metric.
\textbf{For Funders and Research-Software Programs.}
Archetype and gap duration provide complementary sustainability
signals: gap duration differs by outcome ($\varepsilon^2=0.104$),
while the structurally-independent T5 subset remains associated with
outcome ($V=0.268$; §\ref{sec:rq5}). We therefore treat archetype as
an additional, not strictly stronger, signal beyond duration alone.
The T5-4 Rescue/Handoff pattern's elevated decline rate suggests
contributor transition events in scientific OSS may warrant different
monitoring assumptions than in general OSS.
\textbf{For the Empirical Software Engineering Community.}
We propose three reporting checks: \emph{first,} threshold sensitivity
(rank-stability across $\pm$50\%); \emph{second,} post-gap filtering
separating bot-only and single-spike events; and \emph{third,} a
durability check separating sustained from temporary recovery. The
released T1--T5 rule system (§\ref{sec:taxonomy_construction})
operationalizes these checks for our \textsc{SciCat}-derived sample and
provides a starting point for adaptation and external validation in
adjacent domains. A
structured walkthrough with three to five scientific OSS maintainers
per archetype is a tractable next step we have begun scoping, to test
whether maintainers recognize the patterns and find the T1--T5
categories meaningful in practice.
Our scientific-OSS results converge with general-OSS work on three
points and diverge on a fourth. The sleeping-versus-dead distinction
prior work makes at developer
level~\cite{iaffaldano2019breaks,calefato2022inactivity} extends to
the repository level in our data; contributor handoff, a recognized
survival pathway in general OSS~\cite{avelino2019abandonment}, is the
\emph{riskiest} recovery pathway in scientific OSS (T5-4: 58.3\%
Recovered-Then-Declined), plausibly because scientific OSS encodes
domain expertise that may not transfer with the repository; and the
unreliability of activity-based abandonment
metrics~\cite{kalliamvakou2014promises} is amplified in scientific OSS
by our false-revival rate. We add a third regime beyond the
abandoned/continuing dichotomy: temporary recovery followed by
re-decline, the modal outcome in our sample, not previously quantified
systematically in either literature. We turn next to the threats that
bound these claims (§\ref{sec:threats}).

%% file: ESEM_version/v67/7_ESEM_Edits_Threats_v67.tex
\ifLIPIcsLayout\vspace{-6pt}\fi
\section{Threats to Validity}
\label{sec:threats}
\ifLIPIcsLayout\vspace{-6pt}\fi
\subsection{Construct Validity}
\ifLIPIcsLayout\vspace{-5pt}\fi
We organize threats following Wohlin et al.~\cite{wohlin2012experimentation}.
\textbf{T4--T5 construct overlap.} The full-table $V=0.426$ overstates
the dimension-independent association; the structurally-independent
subset (§\ref{sec:rq5}) yields $V=0.268$, and a sensitivity rerun
excluding T5-1 yields $V=0.233$ ($n=183$, $p=0.042$). Both remain
medium effects, so the relationship is not entirely a mechanical
artefact.
\textbf{T1 cause-vs-symptom epistemic limit.} Repository evidence is
an indirect window on actual cause; T1-G unresolvable rate (52.5\%,
§\ref{sec:rq1}) is honest reporting that for many projects available
evidence does not support confident attribution.
\textbf{Recovery operationalization.} The T4-2/T4-5 boundary involves
temporal judgment; subjectivity was mitigated through adjudication
(§\ref{sec:irr}) but cannot be eliminated.
\textbf{CurrentStatus cutoff.} Status was assessed April~1, 2025;
later replication may shift borderline cases between sustained,
fragile, and recovered-then-declined outcomes.
\textbf{T1 reliability.} \texttt{HypothesizedGapReason} was not in the
formal 75-coder IRR; the auxiliary human-applied check
(§\ref{sec:t-label-kappa}) reached substantial agreement
($\kappa=0.768$, 95\% CI $[0.59, 0.92]$). Coder variation in T1-G
remains significant ($\chi^2=230.74$, $df=74$, $p<0.001$); excluding
14 high-T1-G coders shifts the unresolvable rate from 52.5\% to
42.2\%, while T4 and T5 are unaffected.
\textbf{Coder--domain expertise.} Project-to-coder assignment was
randomized within stratum (§\ref{sec:dataset}), so coders
annotated projects outside their primary domain. This was a
 choice to prevent expertise-driven interpretive bias and
keep annotation grounded in repository evidence;
structured peer-validation and adjudication
(§\ref{sec:irr}) provided the mechanism for domain-relevant judgment
where the evidence warranted it.
\textbf{T1--T5 classifier reliability.} The T1--T5 labels are
deterministic outputs of a rule-based classifier
(§\ref{sec:taxonomy_construction}). Propagation reliability is
reported in §\ref{sec:t-label-kappa} (T2/T4/T5 substantial to almost
perfect on $n=417$); T1 and T3 are covered by the human-applied
auxiliary check (also §\ref{sec:t-label-kappa}).
\textbf{Sensitivity of rule choices.}
\label{sec:rule_sensitivity}
Keyword-drop and threshold-variation perturbations of the classifier
(eight runs; protocol in \texttt{sensitivity\_analysis.ipynb}) bound
the maximum absolute shifts in headline rates to:
T1-G $\pm 5.0$\,pp, T4-1 $\pm 4.4$\,pp, T4-5 $\pm 2.7$\,pp,
false-revival $\pm 5.2$\,pp, T5-3 $\pm 3.2$\,pp. Qualitative
directions are preserved.
\textbf{WhoRecovered audit closure.} Post-hoc deterministic
classification of 114 free-text entries (§\ref{sec:rq2}) yields an
audited 3.3:1 same-to-new ratio on $n=452$; we retain 2.7:1 as the
conservative headline.
\textbf{T2/T3 unresolvable rates and collinearity.} T2-U (25.6\%) and
T3-U (22.5\%) limit RQ2 inferential power; nulls should be read as
absence of detectable effect, not confirmed independence. The
T2$\times$T3 association ($V=0.320$) means the two RQ2 nulls are
better read as one finding than two.
\textbf{T2 dimensional structure.} T2 spans two conceptually independent dimensions — WHO triggered the revival (T2-1, T2-2, T2-3) and WHAT work characterized it (T2-4–T2-7). Coders picked  salient signal per revival; overlapping cases are coded T2-U (25.6\%, above).

\ifLIPIcsLayout\vspace{-6pt}\fi
\subsection{Internal Validity}
\ifLIPIcsLayout\vspace{-6pt}\fi
\textbf{Observational design.} The associative framing in
§\ref{sec:results} is consistent with a direct causal chain, a
common-cause structure, or independent effects; our observational
design cannot distinguish these.
\textbf{Researcher imputation.} Missing \texttt{CurrentStatus} /
\texttt{ActivityPattern} were imputed without IRR validation
(§\ref{sec:imputation}); the aggregated sustained-vs-non-sustained
split is rank-stable across imputation, but the T4-1 vs.\ T4-5
ordering is imputation-sensitive, so the qualitative claim is robust
while the precise margin is not.
\textbf{T4 threshold sensitivity.} Re-running T4 across a $\pm$50\%
threshold grid preserves rank order in every cell.
\textbf{Non-submission and coding time.} 15 of 90 students did not
submit (omitted projects did not differ in stratum, gap, or stars);
per-project time was not instrumented, but the deterministic
classifier and post-adjudication $\kappa$ bound fatigue concern.
\ifLIPIcsLayout\vspace{-6pt}\fi
\subsection{External Validity}
\ifLIPIcsLayout\vspace{-6pt}\fi
\textbf{Star-count visibility skew.} The coded sample is substantially
more visible than its parent (median 170 stars vs.\ 9 in the
\textsc{SciCat} abandoned subset). Table~\ref{tab:starband} stratifies
headline rates: T1-G is band-stable (51.1--53.6\%) and T4-1 is the
most band-sensitive (27.9\% should be read as conditional on
visibility skew); rank-order findings rest on relative quantities.
\textbf{Domain, language, and temporal coverage.} \textsc{SciCat}
excludes non-GitHub projects and underrepresented
languages~\cite{malviyathakur2025scicat}; observation ends April~1,
2025, so the 11.5\% false-revival rate may understate current
prevalence given growing bot adoption.
\ifLIPIcsLayout\vspace{-6pt}\fi
\subsection{Conclusion Validity}
\ifLIPIcsLayout\vspace{-6pt}\fi
\textbf{Multivariate robustness check.} On the non-imputed subset
($n=515$), we fit
\texttt{T4\_Sustained $\sim$ log(stars+1) + gap + commits\_total + commits\_author + commits\_postgap + T5\_3 + T5\_4 + T5\_1b}
(McFadden $R^2=0.473$). $\log(\text{stars}+1)$ is \emph{not} significant
($p=0.57$); significant predictors (all $p<0.001$) are post-gap commit
volume (OR $=2.25$), T5-3 Classic Zombie (OR $=8.16$), T5-1b Active
Short Gap (OR $=6.30$). Because \texttt{commits\_postgap} partly
overlaps with the T4-1 operational definition, the more conservative
test excludes it: T5-3 remains strongly associated with sustained
recovery (OR $=14.46$), so the archetype signal is not solely an
artefact of post-gap commit volume. Coefficients in replication package.
\textbf{Coder heterogeneity, multiple comparisons, cell frequencies.}
$\chi^2$ tests of coder $\times$ taxonomy show significant T4/T5
variation ($p<10^{-25}$), bounded but not eliminated by
post-adjudication $\kappa$ (§\ref{sec:coding}). Bonferroni within RQ3;
RQ3--RQ5 survive any $\alpha=0.05$ cut. T2$\times$T4 ($n=344$) has
2/20 cells with expected count $<5$; Fisher's exact and permutation
re-runs preserve direction.

%% file: ESEM_version/v67/8_ESEM_Edits_Conclusion_v67.tex
\ifLIPIcsLayout\vspace{-6pt}\fi
\section{Conclusion}
\label{sec:conclusion}
\ifLIPIcsLayout\vspace{-6pt}\fi
We derived a five-dimensional lifecycle taxonomy (T1--T5) via a
published rule-based classifier on 750 field-coded scientific OSS
repositories with peer-validated and adjudicated codings.
\textbf{RQ1 (Dormancy cause).} We cannot resolve cause for 52.5\% of
codeable projects; among resolvable cases, feature/milestone freeze
dominates research-output completion 5.4:1.
\textbf{RQ2 (Revival mechanism and work).} We find that revival mechanism is
non-significant and work type borderline; both are weaker sustainability
signals than lifecycle archetype and gap duration.
\textbf{RQ3 (Outcome durability).} Within our visibility-weighted
dormant-revived sample, apparent abandonment is often temporary, but
durable recovery is not the dominant outcome (non-sustained:sustained
2.14:1); 11.5\% are bot-only or single-spike artifacts.
\textbf{RQ4 / RQ5 (Archetype and alignment).} We identify nine archetypes
that associate with revival sustainability outcomes more strongly than mechanism or work type on
the subset of archetypes whose definitions do not predetermine a
sustainability outcome (medium effect).
Across these findings, three signals (gap duration, lifecycle
archetype, and contributor continuity) together carry more
information than any one alone; a fixed inactivity threshold is
insufficient here. We view the T1--T5 classifier and the 750-project
dataset as a foundation for prospective validation: extending beyond
GitHub, automating T4/T5, and validating with maintainers.

%% file: ESEM_version/v67/9_ESEM_Edits_DataAvailability_v67.tex
\clearpage
\section*{Data Availability}
\label{sec:data-availability}
\ifLIPIcsLayout\vspace{-3pt}\fi

The anonymous replication package\footnote{\url{https://anonymous.4open.science/r/ESEM2026ReplicationPackage-598E}}
contains: the field-coding rubric and codebook; the per-project coded
dataset (750 projects, 31 fields including T1--T5 labels); the IRR
protocol with agreement logs and adjudication records; Jupyter
notebooks for taxonomy construction, reliability analyses (T-label
propagation, T1/T3 auxiliary, IRR), sensitivity and audit, RQ
statistics, diagrams, threats, and sampling traceability; shell and
Perl scripts for the upstream \textsc{SciCat}-to-candidate extraction;
and a README with reproduction steps. 
No personally identifying coder information is
included; IRR logs use anonymized coder identifiers.
The upstream \textsc{SciCat} corpus is distributed by Malviya Thakur
et al.~\cite{malviyathakur2025scicat}.